\documentclass[usenatbib]{mn2e}
\usepackage{epsfig}
\usepackage{amsmath,amssymb}
\newcommand{\Msol}{{\,\rm M}_\odot}
\newcommand{\Mpc} {{\,\rm Mpc}}
\newcommand{\kpc} {{\,\rm kpc}}
\newcommand{\kms}{{\,\rm {km\,s^{-1}} }}

\title{Velocity distributions in clusters of galaxies}
\author[Faltenbacher and Diemand]
{\parbox[t]\textwidth{Andreas Faltenbacher and Juerg Diemand}
\vspace*{6pt} \\
UCO/Lick Observatory,
University of California at Santa Cruz, 
1156 High Street, Santa Cruz, CA 95064, USA
}
\date{\today}
\begin{document}
\maketitle
\begin{abstract}
We employ a high-resolution dissipationless N-body simulation of a
galaxy cluster to investigate the impact of subhalo selection on the
resulting velocity distributions. Applying a lower limit on the
present bound mass of subhalos leads to high subhalo velocity
dispersions compared to the diffuse dark matter (positive velocity
bias) and to a considerable deviation from a Gaussian velocity
distribution ($kurtosis\sim -0.6$). However, if subhalos are
required to exceed a minimal mass before accretion onto the host, the
velocity bias becomes negligible and the velocity distribution is close to
Gaussian ($kurtosis\sim -0.15$). Recently it has been shown that
the latter criterion results in subhalo samples that agree well with the 
observed number-density profiles of galaxies in clusters. Therefore we
argue that the velocity distributions of galaxies in clusters are
essentially un-biased. The comparison of the galaxy velocity
distribution and the sound speed, derived from scaling relations of
X-ray observations, results in an average Mach number of
1.24. Altogether $65\%$ of the galaxies move supersonically and $8\%$
have Mach numbers larger than 2 with respect to the intra cluster gas. 
\end{abstract}
\section{Introduction}
Velocity distributions in groups and clusters of galaxies can be used
to determine their dynamical masses. It is commonly agreed upon that
line of sight distributions similar to Gaussian reveal relaxed systems
(e.g.~\citealt{Chincarini1977,Halliday2004,Lokas2005}).
Non-gaussianity is usually associated with merging or even
multiple-merging events
(e.g.~\citealt{Colless1996,Cortese2004,Adami2005,Girardi2005}).     
Obviously, only relaxed systems may yield reliable mass
estimates. However, cold dark matter (CDM) simulations 
have revealed high velocity dispersions of subhalos compared to the
diffuse dark matter, even if relaxed systems are considered 
\citep{Ghigna2000,Colin_etal00,Diemand_etal04b}. In other words, the
subhalo populations show a positive velocity bias or are hotter
compared to the diffuse component. For dynamical mass estimates of
observed clusters (e.g.~\citealt{Lokas2005}) it is important to know
whether the velocities of galaxies are biased or not.

In comparing real galaxy clusters with N-body simulations,
subhalos must be associated with galaxies. There have been different 
selection criteria proposed in the literature. As discussed below, the
strength of the velocity bias depends strongly on the subhalo
selection. Therefore we compare the velocity distributions of  
two differently selected subhalo samples derived from the same N-body
cluster. One sample comprises all bound dark matter substructures
above a certain mass limit at present time ($z=0$). This kind of
selection, which has been used in earlier investigations,
leads to a positive velocity bias. However, the 
spatial distribution of these subhalos is less concentrated than the
underlying dark matter distribution and not in agreement with observed
galaxy distributions. \citep{Diemand_etal04b}. A second subhalo
sample, with distributions 
similar to the observed galaxy distributions, contains only those
subhalos, which exceeded a certain mass limit before 
accretion onto the host. Using this accretion time subhalo selection
criterion \cite{Nagai2005} were successful in matching the
distribution of observed cluster galaxies with the results of
N-body simulations. \cite{Conroy2005} use the maximal circular
velocity at the time of accretion to assign luminosities
and stellar masses to (sub)halos and achieve excellent 
agreement of modelled and observed galaxy clustering properties.   
This agreement suggests that the luminosity of a galaxy is related to
the depth of halo potential at the epoch of high starforming activity,
i.e. to its mass or circular velocity before entering the 
group or cluster.  

Semi-analytic modeling of galaxy formation combined with high
resolution dissipationless galaxy cluster simulations
\citep{Springel2001, Gao2004} also produces similar spatial and
velocity distributions for the dark matter and the galaxies. Evidently
the subhalos are populated with galaxies as in 
\cite{Nagai2005} but in a less transparent, more model dependent
way. Similar spatial distributions of galaxies and dark matter were
also found in hydrodynamic cosmological simulations (see
\citealt{Nagai2005,Sommer-Larsen2004,Maccio2005}). 

The spatial distribution of tracers in cosmological dark matter halos
is related to their velocity distribution to a very good approximation
via the spherical, stationary Jeans equation
\citep{Diemand_etal04b,Diemand2005}. A spatially extended component 
(like subhalos) is hotter than the dark matter whereas more
concentrated subsets (like intra cluster light or globular clusters)
are colder. Since cluster galaxies tend to trace the total mass
distribution, one expects little or no difference between galaxy and
dark matter velocity distribution. Clusters of galaxies are formed by
gravitational collapse, which results in a nearly Gaussian velocity
distribution for the diffuse dark component \citep{Hoeft2004,Hansen2006},
thus the velocity distribution of galaxies should be close to Gaussian
as well. In \S\ref{sec:sim} the simulation and the two subhalo samples
are described. In \S\ref{sec:selection} we discuss the impact of the 
selection criterion of the subhalo samples on the velocity
dispersions. Additionally the average Mach number of galaxies orbiting
in the intra cluster medium (ICM) is derived. In \S\ref{sec:summary}
we summarise our results.  
\section{Simulation and subhalo samples}	
\label{sec:sim}
\begin{figure}
\begin{center}
\epsfig{file=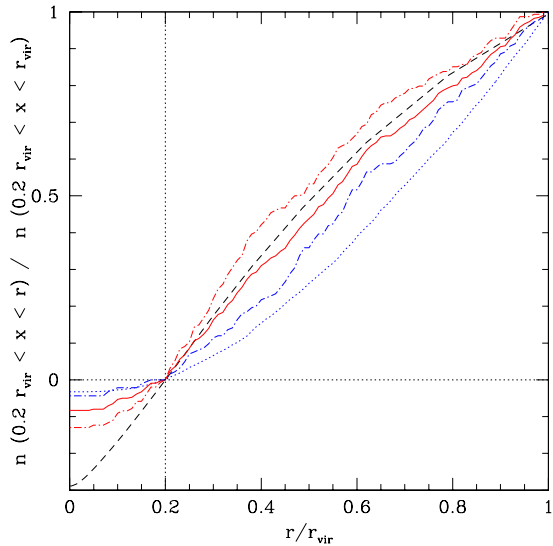,width=0.90\hsize}
\end{center}
\caption{\label{fig1}
Normalised radial number distribution for the total amount of dark
matter (dashed line) and subhalos belonging to the two different samples 
{\bf a} and {\bf b} displayed in blue dotted and red solid lines,
respectively. The dashed-dotted lines display subsamples
of sample {\bf b}. Red and blue colour indicates accretion before and
after $z=0.15$, respectively. The region within $0.2 r_{vir}$ is
excluded from accumulation to reduce the influence of numerical
overmerging.} 
\end{figure}
We analyse a cluster-sized dark matter halo generated within a
cosmological N-body simulation ($\Omega_{m}=0.268$,
$\Omega_{\Lambda}=0.7$, $\sigma_{8}=0.7$, $h_{100} = 0.71$). The peak
mass resolution is $2.2 \times 10^{7}\Msol$ with a softening length of
$1.8\kpc$. The cluster (``D12'' in \citealt{Diemand_etal04b}) has a
virial mass of $3.1 \times 10^{14}\Msol$ at $z=0$ which corresponds to
$\sim14\times 10^{6}$ particles within the virial radius of $1.7\Mpc$ 
\footnote{According to the definition used here the virial radius
encloses 368 times the mean matter density.}.

We create two different subhalo samples. On the one hand we use
SKID\footnote{http://www-hpcc.astro.washington.edu/tools/skid.html}   
with a linking length of $5\kpc$ and identify all bound structures
comprising at least 10 particles as subhalos.  This way we find 4239
subhalos within the virial radius of the cluster at $z=0$. Subsequently,
this subhalo sample is referred to as sample {\bf a}. On the other
hand we trace back the most bound particle of the subhalos, which were
identified by SKID in the same manner as mentioned above, and compare
their positions with those of field halos which are found 
with a friends-of-friends group finder (FOF) in earlier outputs using
a linking length of 0.2 times the mean particle separation. The final
sample encompasses only those 
subhalos, which had progenitors (FOF field halos) containing a minimum
of 200 ($4.4 \times 10^{9}\Msol$) particles at least once during their
field-halo phase. In total, 367 sub-halos meet this criterion. These 
subhalos are assumed to host galaxies. \cite{Nagai2005} and
\cite{Conroy2005} used this approach to assign galaxies to subhalos
derived from pure dark matter N-body simulations. Subsequently, this
sample is referred to as the ``galaxy sample'' or sample {\bf
b}. The galaxy sample is subdivided into two sub-samples according to
accretion times before and after $z=0.15$. The old and the young
sub-samples contain 174 and 193 subhalos, respectively. We do not
intend to assign any stellar properties to the subhalos, however it is
expected that the old sample represents on average redder galaxies,
since star formation in these galaxies may be efficiently suppressed
by interactions with the dense intra cluster medium.   

Figure~\ref{fig1} displays the cumulative radial number distribution
for the different samples and the diffuse dark matter component. The
location of the most bound particle is chosen as centre, which is
assumed to coincide with the central brightest galaxy of observed
clusters. We start the cumulation for all components, subhalos,
galaxies and diffuse dark matter, at $20\%$ of the virial radius ($0.2
r_{vir}$). The inner $20\%$ is excluded because the high 
density environment is likely to artificially remove substructure by
numerical overmerging. Moreover the survival of galaxies in 
this very inner region also depends on the mass distribution
within their inner, baryon dominated parts
\citep{Nagai2005,Maccio2005}. The galaxy sample profile (solid
line) and the diffuse dark matter profile (dashed line) are very
similar, whereas the halos of sample {\bf a} (dotted line) show a
definite  deviation. The splitting of the galaxy sample
according to accretion times, leads to a strongly concentrated old
subsample (red dotted-dashed lines). The young sub-sample (blue
dotted-dashed lines) is more similar to sample  {\bf a}. 

The selection for sample {\bf a} based on the current mass
of the subhalos ignores the history of the individual subhalos. Thus a
recently accreted low mass halo and a tidally striped old (i.e. early
accreted) subhalo are treated equivalently. Due to the steep mass function of
field halos, most systems ever accreted had masses not much larger
than the minimal mass. Those halos that who still lie above this
minimal mass today are mostly systems which have lost little mass,
i.e. recently accreted halos in the outer part of the cluster
(see~\citealt{Kravtsov2004,Zentner2005}). The selection criterion for
the galaxy sample ensures that only subhalos 
with a substantial initial mass (200 particles) are counted as
members of the sample. Since these halos must have retained at least
10 particles to be found by the SKID halo finder at $z=0$, they can
loose $95\%$ of their initial mass on their orbits within the cluster
potential well (or even more if they were more massive at the moment
of accretion). The subhalos of the galaxy sample are durable. In that
respect the galaxy sample is very similar to the diffuse dark matter
component, which by default is indestructible. 
\section{Dependence of the velocity distribution on the subhalo
selection}	 
\label{sec:selection}
\begin{figure}
\begin{center}
\epsfig{file=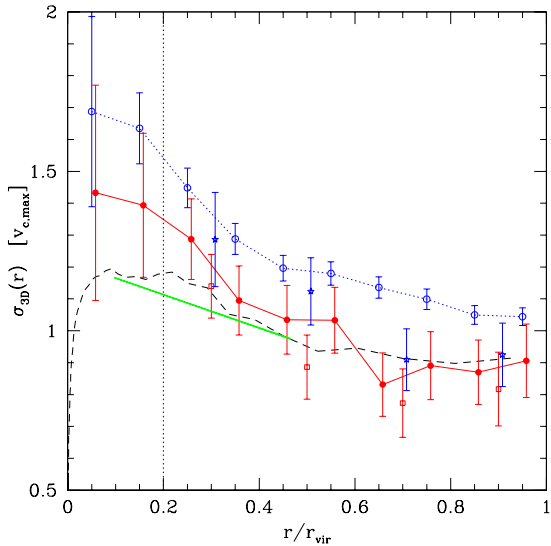,width=0.90\hsize}
\end{center}
\caption{\label{fig2}
Normalised 3D velocity dispersion profiles ($v_{c,max}=958\kms$) for
all dark matter particles (dashed line) and halos belonging to the
subsamples {\bf a} and {\bf b} displayed in dotted and solid lines,
respectively. The error bars display $1\sigma$ deviations. The
region within $0.2 r_{vir}$ is likely to be affected by
overmerging. Open stars and squares indicate the subdivision of 
sample {\bf a} in young and old galaxies, respectively. The solid
green line shows the velocity dispersion derived from X-ray
temperature profiles for clusters of comparable size.}
\end{figure}
Figure~\ref{fig2} displays radially binned velocity dispersions of the
samples {\bf a} and {\bf b} with open and filled symbols, respectively
and the diffuse dark matter indicated by the dashed line. All
dispersions are in units og the circular velocity of the host halo 
$v_{c,max}=958\kms$. The dotted vertical line marks the central region
which is prone to numerical overmerging. Table~\ref{tab:sigma}
displays the characteristic values of the velocity distributions
excluding the inner $0.2 r_{vir}$. The 
qualitative picture does not change if the central volume is included. 
All velocities are computed relative  to the centre of mass velocity
($v_{COM}$), the average velocity of all particles within $r_{vir}$. 
The dispersion of the diffuse dark matter component is shown as black
dashed line. The solid green line shows the velocity dispersion
derived from X-ray temperature profiles for clusters with comparable
size using the relations given in \cite{Vikhlinin2005} and
\cite{Evrard1996}.

For all radii the velocity dispersion of sample {\bf a}
deviates by more than $\sim15\%$ from the diffuse component. The
difference between these two distributions increases towards the
centre, approaching values as large as $\sim30\%$ at $0.2
r_{vir}$. There also appears a slight deviation of the velocity
dispersion of sample {\bf b} compared to the diffuse component,
however, these deviations lie within the $1\sigma$ uncertainty
range. The velocity dispersion profile of the galaxy sample (sample
{\bf b}) and the diffuse dark matter component are very similar.
The old and recently accreted galaxy subsubsamples reveal lower
and higher velocity dispersions, respectively. The velocity
dispersions of the young subsample agree with the total galaxy sample
at large radii but show excess towards the centre. 
Dispersions of the old galaxy subsample deviate below the total galaxy
sample at large radii. The mean velocity dispersions of the subsamples
compared to the velocity dispersions of all galaxy halos are
$\sigma_{old} = 0.95 \sigma_{all}$ and $\sigma_{new} = 1.04
\sigma_{all}$.  

As discussed before, the similarity of the galaxy sample and 
the diffuse component in the density profiles can be explained
by the long lifetime of the sample members. Therefore halos
of sample {\bf b} can be considered as a set of durable 
particles similar to the simulation particles, but with much larger
masses. Due to energy conservation (and without including dynamical
friction or other energy redistributing mechanisms) the gravitational
collapse of a distribution of different mass particles initially leads to
equal velocity dispersions within different mass bins. In this scenario
neither spatial nor velocity biases are expected between sample {\bf b}
and the diffuse component. On the other hand Figure~\ref{fig2}
indicates a prominent velocity bias or offset between the diffuse
component and sample {\bf a}. The average lifetime of sample {\bf a}
is shorter compared to  sample {\bf b}. Sample {\bf a} is weighted
toward subhalos that recently entered the host halo and consequently
move faster. This mechanism shifts the average velocity of the
remaining subhalos in the sample towards higher values and causes a
positive velocity bias (see
\citealt{Ghigna2000,Colin_etal00,Diemand_etal04b}).
\begin{figure}
\begin{center}
\epsfig{file=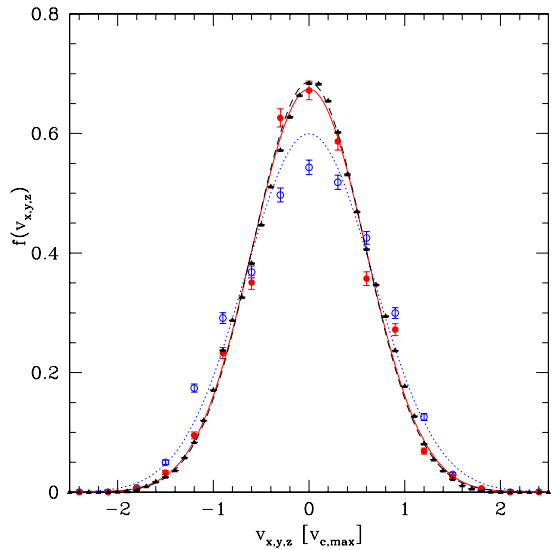,width=0.90\hsize}
\end{center}
\caption{\label{fig3}
Velocity distributions of sample a, b are displayed with blue open and
red filled circles. The diffuse dark matter is indicated by black triangles.
The lines give the Gaussian distribution according to
the mean velocity dispersion of each sample 
Only objects in the range $0.2 < r/r_{vir} < 1$ are used.}   
\end{figure}
\begin{table}
\begin{tabular}{lccc}
$r_{[0.2,1]}$&${\bf a}$&${\bf b}$&${diff}$\\\hline\hline
$\sigma/v_{\rm c,max}$&$ 0.665\pm0.007$&$0.592\pm0.023$&$ 0.5816\pm0.0001$\\
kurtosis              &$ -0.61\pm0.16$ &$ -0.16\pm0.44 $&$-0.241 \pm0.003$\\
$\beta$               &$-0.073\pm0.031$&$  0.17\pm0.11 $&$ 0.1103\pm0.0005$\\
N                     &$  4115        $&$   339        $&$   10.9\times10^6$\\
\end{tabular}
\caption{\label{tab:sigma}
Velocity dispersion, kurtosis ($\bar{v^4} / \sigma^4 - 3$) and anisotropy
parameter $\beta = 1 - 0.5 \sigma^2_{\rm tan} / \sigma^2_{\rm rad}$ 
of the sub-halo sample ({\bf a}), the galaxy sample ({\bf b}) and the
diffuse dark matter ($diff$).}  
\end{table}
Figure~\ref{fig3} compares the projected velocity distributions
of the two subhalo samples and the diffuse component. The dotted,
solid and dashed lines are the Gaussian distributions derived from 
the dispersions of the respective components. Table~\ref{tab:sigma}
displays the actual values of the velocity dispersions in units of the
maximum circular velocity of the host halo ($v_{c,max}=958\kms$). The
distributions of the galaxy sample and the diffuse dark matter are
very similar. The velocity dispersions and kurtosis of these two
samples agree within 1 $\sigma$  uncertainty. The velocity
dispersion of sample {\bf a}, however, exceeds the two others by
$\sim15\%$. The corresponding velocity distribution (open circles, 
Figure~\ref{fig3}) is flat-topped and not well approximated by a
Gaussian distribution (dotted line). There appears to be a lack of
slow moving subhalos and a slight excess of high velocity subhalos 
causing a negative kurtosis of $\sim -0.6$. These features
can naturally be explained by the loss of earlier accreted, slow
moving,  subhalos due to tidal truncation. In this context, loss
means decline in particle numbers below the resolution limit of 10
particles.

The shapes of observed galaxy velocity distributions in relaxed
clusters can in principle be used to infer if they follow a biased,
flat-topped or unbiased, Gaussian distribution. In practice this are
difficult measurements since large numbers of galaxies and careful
removal of interlopers are needed to achieve a significant
result. There are some first hints for flat-topped velocity
distributions: \citet{vanderMarel2000} report a negative  $h_4 =
-0.024\pm0.005$ (which is comparable to the subhalo samples, see
\citealt{Diemand_etal04b}) after stacking 16 CNOC1 clusters and
excluding the cD galaxies. \citet{Lokas2005} found negative kurtosis
values (around $-0.4$) in 5 out of 6 nearby relaxed Abell
clusters. However, the deviations from a Gaussian are only about
1$\sigma$ for these 5 individual systems. Our study suggests that a
kurtosis which is significantly more negative than the one for the
diffuse component (which has $k \sim -0.15$) could be an indicator for
positive velocity bias (and a related spatial anti-bias). It need not
necessarily be related to tangential orbits (negative $\beta$) as
often assumed (see e.g. \citealt{vanderMarel2000} based on models of 
\citealt{Gerhard1993}). 

The orbital anisotropy of the galaxy sample is not significantly
different from the dark matter background, both are slightly radial in
cluster D12. Sample {\bf a} on the other hand shows marginally
tangential orbits. Note that there is significant variation from halo
to halo in the $\beta(r)$ profiles. The average over six relaxed
clusters similar to D12 shows that the dark matter $\beta(r)$ grows
from zero (i.e. isotropic) to about 0.35 near the virial radius and  
total subhalo populations (corresponding to our sample {\bf a}) show a
similar behaviour with a weak tendency to be closer to isotropic on
average \citep{Diemand_etal04b}.

\begin{table}
\begin{center}
\begin{tabular}{ccc}
$M\geq 1$ & $M\geq 2$ & $M\geq 3$ \\\hline\hline
64.44 \% &  8.33 \% & 0.18 \%
\end{tabular}
\caption{\label{tab:mach}
The percentage of galaxies which are expected to exceed the Mach
numbers listed in the upper line.}
\end{center}
\end{table}
The green line in Figure~\ref{fig2} displays the velocity dispersion
(temperature) profile of X-ray gas in clusters with masses comparable
to the cluster analyzed here (see \citealt{Vikhlinin2005}). Despite all
the complex gas physics involved, it is very similar to the
diffuse dark matter and galaxy sample profiles. This finding can be
used to estimate the typical Mach numbers of galaxies with respect 
to the ICM. Assuming adiabatic sound speed ($v_s = \sqrt{\gamma
\sigma_1^2}$, where $\sigma_1$ is the 1D velocity dispersion of the
system and $\gamma=5/3$ is the adiabatic constant) and  integrating
the Gaussian distribution of the galaxies results in an average Mach 
number of $\sim1.24$. The distribution of Mach numbers is displayed in
Table~\ref{tab:mach}. For supersonic galaxy motions leading bow shocks
and ram pressure stripped tails are present
\citep{Stevens1999} which can be detected by X-ray observations. Tails
of supersonic galaxies are expected to be more irregular than 
these of subsonic moving galaxies 
\citep{Roediger2006}. Moreover, since ram pressure is proportional to
the ICM density times the galaxy velocity squared \citep{Gunn1972},
the appearance of leading bow shocks reduces ram pressure and
decreases the stripping efficiency, which may have impact on the
abundance profiles in groups and clusters.
\section{Summary and conclusions}
\label{sec:summary}
Using a cold dark matter simulation of a cluster sized ($3.1 \times
10^{14}\Msol$ at $z=0$) host halo, we find that different subhalo
selection criteria change the resulting velocity distributions.
We analyse the velocity distribution of two differently selected
subhalo samples. Sample {\bf a} contains all presently found subhalos
with masses above $2.2 \times 10^{8}\Msol$ (10 particles). 
Similar selection criteria have commonly been used for investigations 
of the velocity bias in N-body simulations, however they do very
likely not generate subhalo samples which are comparable to galaxies
in groups and clusters. Sample {\bf b} comprises subhalos, which
were able to accumulate more than 200 particles before entering the
host halo. This kind of selection results in number density profiles
which are similar to galaxies observed in groups and clusters (see
\citealt{Kravtsov2004,Nagai2005}). Our main conclusions are as
follows:      
\newline{\bf (1)} In agreement with other authors, we find an enhancement of
the velocity dispersion in the range from $15\%$ to $30\%$ if sample
{\bf a} is compared to the diffuse dark matter component. On average
sample {\bf a} comprises more recently accreted, fast moving,
sub-halos since the early accreted, somewhat more slowly moving, halos
are prone to tidal dissolution. The positive velocity bias in sample
{\bf a} results from a lack of slow moving sub-halos, i.e. a flat
topped non Gaussian velocity distribution with negative kurtosis
$k=-0.6$. 
\newline{\bf (2)} We find no significant velocity bias between sample
{\bf b} and the diffuse component. Both have nearly Gaussian velocity 
distributions ($k \sim -0.15$) and small radial anisotropies
($\beta \sim 0.15$). Since sample {\bf b} resembles the
spatial distributions of galaxies within clusters, it seems reasonable
to identify sample {\bf b} with such galaxies. We conclude, that the
velocity distribution of cluster galaxies is very similar to the 
underlying dark matter velocity distribution. This finding supports
the assumption of not applying a spatial or velocity bias when
estimating the cluster masses via galaxy kinematics (see
e.g.~\citealt{Lokas2005}).
\newline{\bf (3)} The difference between sample {\bf a} and {\bf b}
lies in the lifetimes of the subhalos. Many subhalos of
sample {\bf a} are low mass objects, lying only a little
above the mass limit. If these subhalos lose a small fraction of their
mass due to tidal forces, they will no longer be members of the 
sample. On the other hand members of sample {\bf b} have to lose
at least $95\%$ of their mass to be removed from the sample. Likewise,
massive galaxies in clusters are assumed to  
survive for a long time after accretion. This can explain 
similar properties of the diffuse dark matter, sample {\bf b} and
luminous galaxies in clusters. However, this picture may change in 
smaller and/or older host systems where dynamical friction and tidal
forces are more important. For instance, fossil groups are presumably
old \citep{DOnghia2005} and may have turned a substantial fraction of 
their old, slow moving, satellite galaxies into diffuse intra-group
light \citep{daRocha2005,Faltenbacher2005b}. It is expected, that the
spatial and velocity distributions of fossil groups show similar
features (flattened central number-density profile and a lack 
of slow moving galaxies) as found in sample {\bf a}.
\newline{\bf (4)} The mean velocity dispersions of the whole 
galaxy sample compared to the old galaxy subsample differ by
$5\%$, thus the resulting mass estimates based on $\sigma^2$ would
deviate by a factor of $10\%$. A similar trend is found in
observations, if more recently accreted galaxy populations are
included for the computation of the total velocity dispersion (see
e.g.~\citealt{deOliveira2006}).      
\newline{\bf (5)} We find an average Mach number of
1.24 for galaxies moving within a relaxed cluster
(compare to \citealt{Faltenbacher2005}). Altogether $65\%$ of the 
galaxies move supersonically and $8\%$ show Mach numbers larger 
than 2. The appearance of shocks affects the interaction between
galaxy and cluster gas in various ways. In particular shocks ahead of
supersonic moving galaxies reduces the ram pressure exerted
on the gas in their disks.     
\section*{Acknowledgements}
We are grateful to William G. Mathews for insightful comments on the  
draft of this paper. We thank the anonymous referee who helped us to
improve the original manuscript. AF has been supported by NSF grant
AST 00-98351 and NASA grant NAG5-13275 and JD by the Swiss National
Science Foundation for which we are very thankful.
\end{document}